\documentclass[fleqn,twoside]{article}
\usepackage{espcrc2}

\usepackage{graphicx}
\usepackage[figuresright]{rotating}
\usepackage{psfig}

\newcommand{\AmS}{{\protect\the\textfont2
  A\kern-.1667em\lower.5ex\hbox{M}\kern-.125emS}}

\title{HI Imaging the Low Red-shift Cosmic Web}

\author{Robert Braun\address[MCSD]{ASTRON, \\ 
        P.O. Box 2, 7990 AA Dwingeloo, The Netherlands}%
        }
       
\begin{document}

\begin{abstract}

Only in recent years has the realization emerged that galaxies do not dominate
the universal baryon budget but are merely the brightest pearls of an
underlying Cosmic Web. Although the gas in these inter-galactic filaments is
moderately to highly ionized, QSO absorption lines have shown that the surface
area increases dramatically in going down to lower HI column densities. The
first image of the Cosmic Web in HI emission has just been made of the Local
Group filament connecting M31 and M33. The corresponding HI distribution
function is in very good agreement with that of the QSO absorption lines,
confirming the 30-fold increase in surface area expected between 10$^{19}$
cm$^{-2}$ and 10$^{17}$ cm$^{-2}$. The critical observational challenge is
crossing the ``HI desert'', the range of log(N$_{HI}$) from about 19.5 down to
18, over which photo-ionization by the intergalactic radiation field produces
an exponential decline in the neutral fraction from essentially unity down to
a few percent. Nature is kinder again to the HI observer below
log(N$_{HI}$)~=~18, where the neutral fraction decreases only very slowly with
log(N$_{HI}$). With the SKA we can begin the systematic study of the Cosmic
Web beyond the Local Group. With moderate integration times, the necessary
resolution and sensitivity can be achieved out to distances beyond the Virgo
cluster. When combined with targeted optical and UV absorption line
observations, the total baryonic masses and enrichment histories of the Cosmic
Web will be determined over the complete range of environmental
over-densities.

\vspace{1pc}
\end{abstract}

\maketitle

\section{Introduction}

Extragalactic astronomy has traditionally focused on the regions of extreme
cosmic over-density that we know as galaxies. Only in recent years has the
realization emerged that galaxies do not dominate the universal baryon budget
but are merely the brightest pearls of an underlying Cosmic Web.  Filamentary
components extending between the massive galaxies are a conspicuous prediction
of high resolution numerical models of structure formation (eg. Dav\'e et
al. \cite{dave99}, \cite{dave01}). Such calculations suggest that in the
current epoch, cosmic baryons are almost equally distributed by mass amongst
three components: (1) galactic concentrations, (2) a warm-hot intergalactic
medium (WHIM) and (3) a diffuse intergalactic medium. These three components
are each coupled to a decreasing range of baryonic over-density:
$log(\rho_{\sc H}/\overline \rho_{\sc H})>3.5$, 1--3.5, and $<$ 1 and are
probed by QSO absorption lines with specific ranges of neutral column density:
$log(N_{HI})~>~18$, 14--18, and $<$ 14. The neutral fraction is thought to
decrease with decreasing column density from about 1\% at log(N$_{HI}$)~=~17,
to less than 0.1\% at log(N$_{HI}$)~=~13. Although a very wide range of
physical conditions can be found within galaxies, the WHIM is thought to be a
condensed shock-heated phase with temperature in the range 10$^5$--10$^7$~K,
while the diffuse IGM is predominantly photo-ionized with temperature near
10$^4$~K.

The strongest observational constraints on this picture come from the
statistics of the QSO absorption lines.  Enough of such QSO spectra have been
obtained to allow good statistical determinations to be made of the rate of
occurrence of intervening absorbers as function of their column density. By
binning such data in red-shift intervals, it has even been possible to gauge
the cosmic evolution of intervening absorbers (\cite{stor00}). Inter-galactic
space has apparently become continuously tidier by about an order of magnitude
from red-shifts of several down to zero; with a decreasing cross-section of
high column absorbers. At the current epoch we can now confidently predict
that in going down from HI column densities of 10$^{19}$ cm$^{-2}$ (which
define the current ``edges'' of well-studied nearby galaxies in HI emission)
to 10$^{17}$ cm$^{-2}$, the surface area will increase by a factor of 30. The
critical observational challenge is crossing the ``HI desert'', the range of
log(N$_{HI}$) from about 19.5 down to 18, over which photo-ionization by the
intergalactic radiation field produces an exponential decline in the neutral
fraction from essentially unity down to a few percent (eg. Dove \& Shull
\cite{dove94}). Nature is kinder again to the HI observer below
log(N$_{HI}$)~=~18, where the neutral fraction decreases only very slowly with
log(N$_{HI}$). The baryonic mass traced by this gas (with a 1\% or less
neutral fraction) is expected to be comparable to that within the galaxies, as
noted above.

\begin{figure}[htb]
\resizebox{\hsize}{!}{\includegraphics{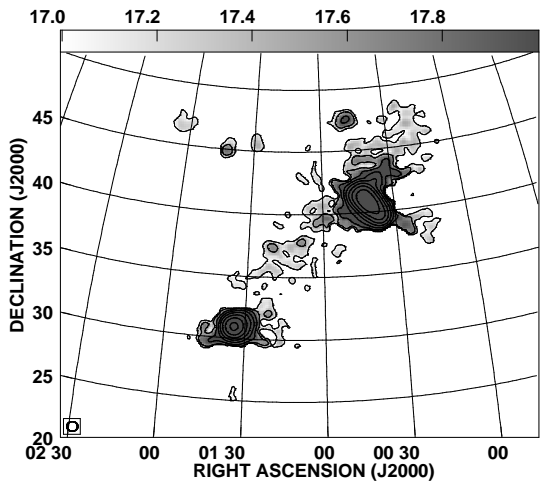}}
 \caption{Integrated HI emission from features which are kinematically
 associated with M31 and M33.  The grey-scale varies between
 log(N$_{HI}$)~=~17~--~18, for N$_{HI}$ in units of cm$^{-2}$. Contours
 are drawn at log(N$_{HI}$)~=~17, 17.5, 18, $\dots$ 20.5. M31 is located at
 (RA,Dec)~=~(00:43,+41$^\circ$) and M33 at (RA,Dec)~=~(01:34,+30$^\circ$), The
 two galaxies are connected by a diffuse filament joining the systemic
 velocities.}
\label{fig:m31m33}
\end{figure}

But how are these diffuse systems distributed and what are their kinematics?
These are questions which can not be addressed with the QSO absorption line
data. The areal density of suitable background sources is far too low to allow
``imaging'' of the intervening low column density systems in
absorption. Direct detection of the free-free continuum or recombination line
emission from the ionized gas has also proven well beyond the capabilties of
current X-ray and optical instrumentation. For example the expected H$\alpha$
emission measure is only about EM~=~5$\times10^{-4}$ cm$^{-6}$ pc. The very
best current H$\alpha$ imaging results reach down to about EM~=~0.1 cm$^{-6}$
pc, which is still orders of magnitude removed from what would be needed.

\begin{figure}[htb]
\resizebox{\hsize}{!}{\includegraphics{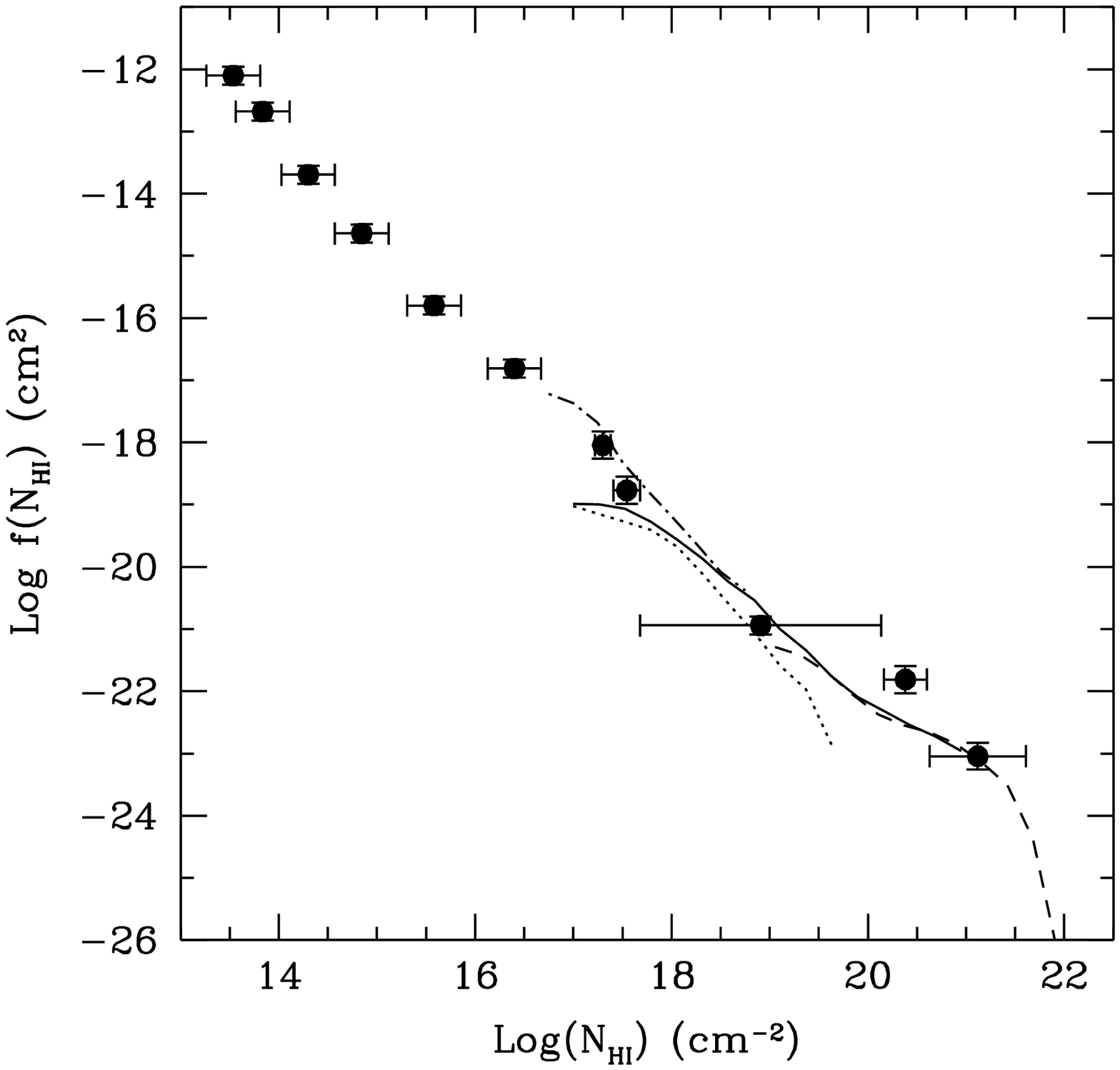}}
 \caption{The distribution function of HI column density
 due to M31 and it's environment. The data from three HI surveys of
 M31 are combined in this figure to probe column densities over a
 total range of some five orders of magnitude. The dashed line is from
 the WSRT mosaic (Braun et al. \cite{brau04b}) with 1$^\prime$
 resolution over 80$\times$40~kpc, the dotted and solid lines from our
 GBT survey (Thilker et al. \cite{thil04}) with 9$^\prime$ resolution
 over 95$\times$95~kpc and the dot-dash line from the wide-field WSRT survey
 (Braun \& Thilker \cite{brau04}) with
 48$^\prime$ resolution out to 150~kpc radius. The filled circles with
 errorbars are the low red-shift QSO absorption line data as tabulated
 by Corbelli \& Bandiera (\cite{corb02}). }
\label{fig:fnhi}
\end{figure}

\section{The First Cosmic Web Images}

Although conventional imaging in the 21cm emission line of neutral hydrogen
has not typically reached column densities below about 10$^{19}$ cm$^{-2}$,
this is not a fundamental limitation. Long integrations with an
(almost-)filled aperture can achieve the required brightness sensitivity to
permit direct imaging of the small neutral fraction within the Cosmic Web
filaments between galaxies. The first detection of such diffuse filaments in
the extended environment of M31 has just been made by Braun \& Thilker
(\cite{brau04}). This was accomplished by utilizing total power measurements
made with the fourteen 25m dishes of the Westerbork Synthesis Radio Telescope
(WSRT). A series of drift-scan observations were used to obtain
Nyquist-sampled HI imaging of a region 60$\times$30 degrees region in extent,
centered approximately on the position of M31.  Although the angular
resolution is low (effective beam of 49~arcmin, corresponding to 11~kpc at the
M31 distance) the column density sensitivity is very high (4$\times10^{16}$
cm$^{-2}$ rms over 17~km~s$^{-1}$).  A diffuse filament is detected connecting
the systemic velocities of M31 to M33 (at a projected seperation of 200 kpc)
and also extending away from M31 in the anti-M33 direction as shown in
Figure~1. This diffuse filament appears to be fueling denser gaseous streams
and filaments in the outskirts of both galaxies.  Peak neutral column
densities within the filament only amount to some 3$\times10^{17}$
cm$^{-2}$. The extremely diffuse nature of the HI has been confirmed by
pointed Green Bank Telescope (GBT) observations of a local peak in the
filament which yields the same low peak column density ($3\times10^{17}$
cm$^{-2}$), despite a telescope beam area that is 25 times smaller.

The interaction zone of the diffuse filament with M31 has been studied in
complimentary surveys: a 6$\times$6 degree field imaged with the GBT (Thilker
et al {\cite{thil04}) and a 5$\times$2 degree WSRT mosaic of nearly 200
synthesis pointings (Braun et al {\cite{brau02},\cite{brau04b}). Our three
surveys permit calculation of the HI distribution function from HI emission
measurements (rather than QSO absorption measurements) over an unprecedented
range in log(N$_{HI}$)~=~17.2 to log(N$_{HI}$)~=~21.9 as shown in
Figure~2. The N$_{HI}$ data for the M31 environment were normalized to the
average space density of galaxies using the HIMF of Zwaan et
al. (\cite{zwaa03}).

The HI distribution function of these structures agrees very well with that of
the low red-shift QSO absorption lines which are also plotted in the figure as
filled circles with error bars.  The predicted factor of 30 increase in
surface covering factor for low N$_{HI}$ emission has been observationally
verified. In so doing, it has been possible to provide the first 
image of a Lyman Limit absorption system. The morphology and kinematics are
fully in keeping with the cosmic web hypothesis outlined above. We are now in
a position to witness the continuing gaseous fueling of normal galaxies with
direct imaging.

\section{The SKA and the Cosmic Web}

The exquisite sensitivity of the SKA will permit the direct kinematic imaging
of the Cosmic Web in a broad range of galaxy environments. Tapering of the
total collecting area to an angular resolution of about 150'' will provide a
brightness sensitivity of 8~mK over 1~km~s$^{-1}$ in an hour of integration. A
four hour integration will then suffice to achieve an HI column density
sensitivity of $3.2\times10^{16}$ cm$^{-2}$ rms over the typical 20~km~s$^{-1}$
linewidth. The 150'' beamsize corresponds to a physically interesting linear 
resolution of better than 15~kpc out to distances of 21~Mpc. The extended
environments of thousands of galaxies; be they isolated, in loose groups or
even in the Virgo cluster are all within reach. 

As a specific example of the type of study that will be possible, consider a
blind fully-sampled survey of a portion of the super-galactic plane filament
between RA~=~12.2 and 12.9 hours and Dec.~=~+2 and +10$^\circ$. More than 130
galaxies with associated HI, covering the full range of Hubble types, are
currently known in this region at distances lying between the Local Group and
the Virgo cluster. A complete imaging survey of this region can be obtained
with some 80 pointings (of a 1.1 square degree instantaneous FOV) reaching a
column density depth of $3.2\times10^{16}$ cm$^{-2}$ rms over 20~km~s$^{-1}$ in
300 hours of integration. Determination of gaseous fueling rates will likely
need to be estimated statistically, since only the radial (and not the
transverse) velocity components can be directly measured. Hence the need for
large samples of galaxies. Once such an imaging study has been obtained, a
complimentary directed study can be carried out of optical and UV absorption
toward suitably located background QSO's behind individual filaments. In this
way the metallicity and ionization state of the gas can be established. With
this information, total baryonic masses and enrichment histories of the Cosmic
Web can be determined over the complete range of environmental over-density.


\begin{thebibliography}{9}

\bibitem{brau04} Braun, R., Thilker, D.\,A., 2004 A\&A, 417, 421.
\bibitem{brau02} Braun, R., Thilker, D.\,A., Corbelli, E., Walterbos,
  R.\,A.\,M., 2002, http://www.astron.nl/newsletter/2002-1/index.html 
\bibitem{brau04b} Braun, R., Thilker, D.\,A., Corbelli, E., Walterbos,
  R.\,A.\,M., 2004, in prep.
\bibitem{corb02} Corbelli, E., Bandiera, R., 2002, ApJ, 567, 712.
\bibitem{dave99} Dav\'e, R., Hernquist, L., Katz, N., Weinberg, D.\, H., 1999,
  ApJ, 511, 521. 
\bibitem{dave01} Dav\'e, R., Cen, R., Ostriker, J.\,P., et al., 2001, ApJ,
  552, 473. 
\bibitem{dove94} Dove, J.\,B., Shull, J.\,M., 1994, ApJ 423, 196.
\bibitem{stor00} Storrie-Lombardi, L.\,J., 2000, ApJ 543, 552.
\bibitem{thil04} Thilker, D.\,A., Braun, R., Walterbos, R.\,A.\,M., et
  al. 2004 ApJ, 601, L39.
\bibitem{zwaa03} Zwaan, M.\,A., Staveley--Smith L.,  Koribalski,
  B.\,S., et al. 2003, AJ, 125, 2842.

\end{thebibliography}
\end{document}